\documentclass[a4paper]{llncs}

\usepackage[english]{babel}
\usepackage{csquotes}
\usepackage{amsmath,amssymb}
\usepackage{graphicx}
\usepackage{enumerate}
\usepackage{verbatim}

\usepackage[style=alphabetic,citestyle=alphabetic,sorting=nty,firstinits=true,backend=bibtex]{biblatex}
\bibliography{bibliography2.bib}{}
\DeclareBibliographyCategory{excluded}

\usepackage[usenames,dvipsnames]{xcolor}
\usepackage{mathtools}
\usepackage{hyperref}

\newcommand{\env}[2]{\begin{#1}#2\end{#1}}%
\newcommand{\envparam}[3][]{\begin{#2}[#1]#3\end{#2}}%

\spnewtheorem*{convention}{Convention}{\bfseries\itshape}{}

\def \N {\mathbb{N}}
\def \Z {\mathbb{Z}}
\def \Q {\mathbb{Q}}
\def \R {\mathbb{R}}
\def \C {\mathbb{C}}

\def \L {\mathcal{L}}
\def \u {{\vec{u}}}
\def \v {{\vec{v}}}

\DeclarePairedDelimiter\abs{\lvert}{\rvert}
\DeclareMathOperator{\supp}{supp}
\DeclareMathOperator{\Ann}{Ann}
\DeclareMathOperator{\bbox}{Box}
\DeclareMathOperator{\opc}{ord}%
\newcommand{\makeinner}[2]{ \, #1 \mid #2 \, }
\newcommand{\makeset}[2]{\{ \makeinner{#1}{#2} \}}
\newcommand{\makesetbig}[2]{\big\{ \, #1 \; \big| \; #2 \, \big\}}%
\newcommand{\gen}[1]{\langle #1 \rangle}

\title{An Algebraic Geometric Approach to Nivat's Conjecture}
\author{Jarkko Kari \and Michal Szabados}
\institute{
    Department of Mathematics and Statistics,\\
    University of Turku, 20014 Turku, Finland
}

\begin{document}
\maketitle

\env{abstract}{We study  multidimensional configurations (infinite words) and subshifts of low pattern complexity using tools of algebraic geometry.
We express the configuration as a multivariate formal power series over integers and investigate the setup when there is a non-trivial annihilating polynomial: a non-zero polynomial whose formal product with the power series is zero. Such annihilator exists, for example, if the number of
distinct patterns of some finite shape $D$ in the configuration
is at most the size $|D|$ of the shape. This is our low pattern complexity assumption.
We prove that the configuration must be a sum of periodic configurations over integers, possibly with unbounded values. As a specific application of the method we obtain an asymptotic version of the well-known Nivat's conjecture: we prove that any two-dimensional, non-periodic configuration can satisfy the low pattern complexity assumption with respect to only finitely many distinct rectangular shapes $D$.
}


\section{Introduction}

Consider configuration $c\in A^{\Z^d}$,
a $d$-dimensional infinite array filled by symbols from finite alphabet $A$. Suppose that
for some finite observation window
$D\subseteq \Z^d$, the number of distinct patterns of shape $D$ that exist in $c$ is small, at most the cardinality $|D|$ of $D$. We investigate global regularities and structures in $c$ that are enforced by such local complexity assumption.

Let us be more precise on the involved concepts. As usual, we denote by $c_\v\in A$ the symbol in $c$ in position $\v\in\Z^d$.
For $\u\in\Z^d$, we say that $c$ is {\em $\u$-periodic\/} if $c_\v=c_{\v+\u}$ holds for all $\v\in\Z^d$,
and $c$ is {\em periodic\/} if it is $\u$-periodic for some $\u\neq 0$.
For a finite domain $D\subseteq \Z^d$, the
elements of $A^D$ are {\em $D$-patterns}. For a fixed $D$,
we denote by $c_{\v+D}$ the $D$-pattern in $c$ in position $\v$,
that is, the pattern $\u\mapsto c_{\v+\u}$ for all $\u\in D$.
The number of distinct $D$-patterns in $c$ is the {\em $D$-pattern complexity\/} $P_c(D)$ of $c$. Our assumption of
low local complexity is
\begin{equation}
\label{eq:lowcomplexity}
P_c(D)\leq |D|,
\end{equation}
for some finite $D$.

\subsection*{Nivat's conjecture}

There are specific examples in the literature of open problems in this framework. {\em Nivat's conjecture\/} (proposed by M. Nivat in his keynote address in ICALP 1997~\cite{Nivat97})
claims that in the two-dimensional case $d=2$, the low
complexity assumption (\ref{eq:lowcomplexity}) for a rectangle $D$ implies that $c$ is periodic.
The conjecture is a natural generalization of the one-dimensional Morse-Hedlund theorem that states that if a bi-infinite word contains at most $n$ distinct subwords of length $n$ then the word must be periodic~\cite{MorseHedlund38}.
In the two-dimensional setting for $m,n\in\N$ we denote by
$P_c(m,n)$ the complexity $P_c(D)$ for the $m\times n$ rectangle $D$.

\begin{conjecture}[Nivat's conjecture]
If for some $m,n$ we have $P_c(m,n)\leq mn$ then $c$ is periodic.
\end{conjecture}

The conjecture has recently raised wide interest, but it remains unsolved.
In~\cite{EpifanioKoskasMignosi03}  it was shown $P_c(m,n)\leq mn/144$ is enough to guarantee the periodicity of $c$. This bound was improved  to $P_c(m,n)\leq mn/16$ in~\cite{QuasZamboni04}, and
recently to $P_c(m,n)\leq mn/2$ in~\cite{CyrKra13x}. Also the cases of narrow rectangles have been investigated: it was shown in~\cite{SanderTijdeman02} and recently in~\cite{CyrKra13y} that $P_c(2,n)\leq 2n$ and $P_c(3,n)\leq 3n$, respectively, imply that $c$ is periodic.

The analogous conjecture in the higher dimensional setups $d\geq 3$ is false~\cite{SanderTijdeman00}. The following example recalls a simple counter example for $d=3$.

\begin{example}
\label{ex:3d}
Fix $n\geq 3$, and consider the following $c\in \{0,1\}^{\Z^3}$ consisting of two perpendicular lines of $1$'s on a $0$-background, at distance $n$ from each other: $c(i,0,0)=c(0,i,n)=1$ for all $i\in \Z$, and $c(i,j,k)=0$ otherwise. For $D$ equal to the $n\times n\times n$  cube we
have $P_c(D)=2n^2+1$ since the $D$-patterns in $c$ have at most a single $1$-line piercing a face of the cube. Clearly $c$ is not periodic although
$P_c(D)=2n^2+1<n^3=|D|$. Notice that $c$ is a ``sum'' of two periodic components (the lines of $1$'s). Our results imply that any counter example must decompose into a sum of periodic components.
\qed
\end{example}

\subsection*{Periodic tiling problem}

Another related open problem is the {\em periodic (cluster) tiling problem\/} by Lagarias and Wang~\cite{LagariasWang}.
A (cluster) tile is a finite  $D\subset \Z^d$. Its co-tiler is any subset $C\subseteq \Z^d$ such that
\begin{equation}
\label{eq:tiling}
D\oplus C = \Z^d.
\end{equation}
The co-tiler can be interpreted as the set of positions where copies of $D$ are placed so that they together cover the entire $\Z^d$ without overlaps. Note that the tile $D$ does not need to be connected -- hence the term
``cluster tile'' is sometimes used. The tiling is by translations of $D$ only: the tiles may not be rotated.

It is natural to interpret any $C\subseteq \Z^d$ as the binary
configuration $c\in \{0,1\}^{\Z^d}$ with $c_\v=1$ if and only if $\v\in C$.
Then the tiling condition (\ref{eq:tiling}) states that $C$ is a co-tiler for $D$ if and only if
the ($-D$)-patterns in the corresponding configuration $c$ contain exactly a single $1$ in the background of $0$'s. In fact, as co-tilers of $D$ and $-D$ coincide~\cite{Szegedy}, this is equivalent to all $D$-patterns having a single $1$.

We see that the set ${\cal C}$ of all co-tiler configurations
for $D$ is a {\em subshift of finite type\/}~\cite{LindMarcus}.
We also see that the low local complexity assumption (\ref{eq:lowcomplexity}) is satisfied. We even have $P_{\cal C}(D)\leq |D|$ where we denote by $P_{\cal C}(D)$ the number of distinct
$D$-patterns found in the elements of the subshift ${\cal C}$.

\begin{conjecture}[Periodic Tiling Problem]
If tile $D$ has a co-tiler then it has a periodic co-tiler.
\end{conjecture}

This conjecture was first formulated in~\cite{LagariasWang}. In the one-dimensional case it is easily seen true, but already for $d=2$
it is open. Interestingly, it is known that if $|D|$ is a prime number then {\it every\/} co-tiler of $D$ is periodic~\cite{Szegedy} (see also our \autoref{ex:prime}).
The same is true if $D$ is connected, that is, a polyomino~\cite{BeauquierNivat}.

\subsection*{Our contributions}

We approach these problems using tools of algebraic geometry. Assuming alphabet
$A\subseteq\Z$,
we express configuration
$c$ as a formal power series over $d$ variables and with coefficients in $A$.
The complexity assumption (\ref{eq:lowcomplexity}) implies that there is a non-trivial polynomial that annihilates the power series under formal multiplication (\autoref{lem-low-cplx-has-annihilator}). This naturally leads to the study of the annihilator ideal of the power series, containing all the polynomials that annihilate it. Using Hilbert's Nullstellensatz we prove that the ideal contains polynomials of particularly simple form (\autoref{thm-product-of-differences}). In particular, this implies that $c=c_1+\dots +c_m$ for some periodic $c_1,\dots ,c_m$ (\autoref{thm-decomposition-theorem}). This decomposition result is already an interesting global structure on $c$, but to prove periodicity we would need $m=1$.

We study the structure of the annihilator ideal in the two-dimensional setup, and prove that it is always a radical (\autoref{lem-ann-is-radical}). This leads to a stronger decomposition theorem (\autoref{thm-strong-decomposition}). In the case of Nivat's conjecture we then provide an asymptotic result (\autoref{thm-main-result}): for any non-periodic configuration $c$ there are only finitely many pairs $m,n\in\N$ such that $P_c(m,n)\leq mn$.

Due to the strict page limit the proofs in the latter part of the paper are omitted.

\section{Basic Concepts and Notation}


For a domain $R$ -- which will usually be the whole numbers $\Z$ or complex numbers $\C$ -- denote by $R[x_1,\dots,x_d]$ the set of polynomials over $R$ in $d$ variables. We adopt the usual simplified notation: for a $d$-tuple of non-negative integers $\v = (v_1,\dots,v_d)$ set $X^\v = x_1^{v_1}\dots x_d^{v_d}$, then we write
\env{align*}{
    R[X] = R[x_1,\dots,x_d]
}
and a general polynomial $f \in R[X]$ can be expressed as $f = \sum a_\v X^\v$, where $a_\v \in R$ and the sum goes over finitely many $d$-tuples of non-negative integers $\v$. If we allow $\v$ to contain also negative integers we obtain \emph{Laurent polynomials}, which are denoted by $R[X^{\pm1}]$. Finally, by relaxing the requirement to have only finitely many $a_\v \ne 0$ we get \emph{formal power series}:
\env{align*}{
    R[[X^{\pm1}]] = \makesetbig{\!\sum a_\v X^\v}{\v \in \Z^d,\ a_\v \in R}.
}
Note that we allow negative exponents in formal power series.

Let $d$ be a positive integer. Let us define a $d$-dimensional \emph{configuration} to be any formal power series $c \in \C[[X^{\pm1}]]$:
\env{align*}{
    c = \sum_{\v \in \Z^d} c_\v X^\v.
}
A configuration is \emph{integral} if all coefficients $c_\v$ are integers, and it is \emph{finitary} if there are only finitely many distinct coefficients $c_\v$. In the case the coefficients are not given explicitly we denote the coefficient at position $\v$ by a subscript.

Classically in symbolic dynamics configurations are understood as elements of $A^{\Z^d}$.
Because the actual names of the symbols in the alphabet $A$ do not matter, they can be chosen to be integers. Then such a ``classical'' configuration can be identified with a finitary integral configuration by simply setting the coefficient $c_\v$ to be the symbol at position $\v$.

The first advantage of using formal power series is that a multiplication by a Laurent polynomial is well defined and results again in formal power series. For example, $X^\v c$ is a translation of $c$ by the vector $\v$. Another important example is that $c$ is periodic if and only if there is a non-zero $\v \in \Z^d$ such that $(X^\v-1)c = 0$. Here the right side is understood as a constant zero configuration.

For a polynomial $f(X) = \sum a_\v X^\v$ and a positive integer $n$ define $f(X^n) = \sum a_\v X^{n\v}$. The following example, and the proof of Lemma~\ref{lem-expanding-annihilator}, use the well known fact that
for any integral polynomial $f$ and prime number $p$, we have $f^p(X) \equiv f(X^p) \pmod{p}$.

\begin{example}
\label{ex:prime}
Our first example concerns the periodic tiling problem. We
provide a short proof of the fact -- originally proved in~\cite{Szegedy} -- that if the size $p=|D|$ of tile $D$ is a prime number then all co-tilers $C$ are periodic. When the tile $D$
is represented as the Laurent polynomial $f(X)=\sum_{\v \in D} X^\v$ and the co-tiler $C$
as the power series $c(X)=\sum_{\v \in C} X^\v$, the tiling condition (\ref{eq:tiling})
states that $f(X)c(X)=\sum_{\v \in \Z^d} X^\v$. Multiplying both sides by $f^{p-1}(X)$, we get
$$f^p(X)c(X)=\sum_{\v \in \Z^d} p^{p-1}X^\v\equiv 0\pmod{p}.$$ On the other hand, since $p$ is a prime, $f^p(X) \equiv f(X^p) \pmod{p}$ so that $$f(X^p)c(X)\equiv 0\pmod{p}.$$
Let $\v\in D$ and $\vec{w}\in C$ be arbitrary. We have
$$0\equiv [f(X^p)c(X)]_{\vec{w}+p\v} = \sum_{\u\in D} c(X)_{\vec{w}+p\v-p\u}\pmod{p}.$$
The last sum is a sum of $p$ numbers, each $0$ or $1$, among which there is at least one $1$ (corresponding to $\u=\v$). The only way for the sum to be divisible by $p$  is by having each summand equal to $1$. We have that $\vec{w}+p(\v-\u)$ is in $C$ for all $\u,\v\in D$ and $\vec{w}\in C$, which means
that $C$ is $p(\v-\u)$-periodic for all $\u,\v\in D$.\qed

\end{example}

The next lemma grants us that for low complexity configurations there exists at least one
Laurent polynomial that annihilates the configuration by formal multiplication.

\env{lemma}{
    \label{lem-low-cplx-has-annihilator}
    Let $c$ be a configuration and $D \subset \Z^d$ a finite domain such that $P_c(D) \leq |D|$. Then there exists a non-zero Laurent polynomial $f \in \C[X^{\pm1}]$ such that $fc=0$.
}
\env{proof}{
    Denote $D = \{\vec{u_1},\dots,\vec{u_n}\}$ and consider the set
    \env{align*}{
        \makeset{(1,c_{\vec{u_1}+\v},\dots,c_{\vec{u_n}+\v})}{\v \in \Z^d}.
    }
    It is a set of complex vectors of dimension $n + 1$, and because $c$ has low complexity there is at most $n = \abs D$ of them. Therefore there exists a common orthogonal vector $(a_0,\dots,a_n)$. Let $g(X) = a_1X^{-\vec{u_1}} + \dots + a_nX^{-\vec{u_n}}$, then the coefficient of $gc$ at position $\v$ is
    \env{align*}{
        (gc)_\v = a_1 c_{\u_1+\v} + \dots + a_n c_{\vec{u_n}+\v} = -a_0,
    }
    that is, $gc$ is a constant configuration. Now it suffices to set $f = (X^\v-1)g$ for arbitrary non-zero vector $\v \in \Z^d$.
\qed}

\section{Annihilating Polynomials and Decomposition Theorem}
\label{sec-annihilators}



\autoref{lem-low-cplx-has-annihilator} motivates the following definitions. Let $c$ be a configuration. We say that a Laurent polynomial $f$ \emph{annihilates} (or is an \emph{annihilator} of) the configuration if $fc=0$. Define
\env{align*}{
    \Ann(c) = \makesetbig{f \in \C[X]}{fc = 0}.
}
It is the set of all annihilators of $c$. Clearly it is an ideal of $\C[X]$. The zero polynomial annihilates every configuration; let us call an annihilator \emph{non-trivial} if it is non-zero. Note that the configuration is periodic if and only if $X^\v - 1 \in \Ann(c)$ for some non-zero $\v \in \Z^d$.

We defined $\Ann(c)$ to consist of complex polynomials, so that we can later use Hilbert's Nullstellensatz directly, as it requires polynomial ideals over algebraically closed field. We shall however occasionally work with integer coefficients and Laurent polynomials when it is more convenient.

Recall that in the case of Nivat's conjecture and Periodic tiling problem we study finitary integral configurations, which by \autoref{lem-low-cplx-has-annihilator} have a non-trivial annihilator. Moreover there is an integer annihilating polynomial -- actually for integral configurations $\Ann(c)$ is always generated by integer polynomials. 

If $Z = (z_1, \dots, z_d) \in \C^d$ is a complex vector then it can be plugged into a polynomial. In particular, plugging into a monomial $X^\v$ results in $Z^\v = z_1^{v_1} \cdots z_d^{v_d}$.

\env{lemma}{
    \label{lem-expanding-annihilator}
    Let $c(X)$ be a finitary integral configuration and $f(X) \in \Ann(c)$ a non-zero integer polynomial. Then there exists an integer $r$ such that for every positive integer $n$ relatively prime to $r$ we have $f(X^n) \in \Ann(c)$.
}
\env{proof}{
    Denote $f(X) = \sum a_\v X^\v$. First we prove the claim for the case when $n$ is a large enough prime.

    Let $p$ be a prime, then we have $f^p(X) \equiv f(X^p) \pmod{p}$. Because $f$ annihilates $c$, multiplying both sides by $c(X)$ results in
    \env{align*}{
        0 \equiv f(X^p)c(X) \pmod{p}.
    }
    The coefficients in $f(X^p)c(X)$ are bounded in absolute value by
    \env{align*}{
        s = c_{max}\sum \abs {a_\v},
    }
    where $c_{max}$ is the maximum absolute value of coefficients in $c$. Therefore if $p > s$ we have $f(X^p)c(X) = 0$.

    For the general case, set $r = s!$. Now every $n$ relatively prime to $r$ is of the form $p_1\cdots p_k$ where each $p_i$ is a prime greater than $s$. Note that we can repeat the argument with the same bound $s$ also for polynomials $f(X^m)$ for arbitrary $m$ -- the bound $s$ depends only on $c$ and the (multi)set of coefficients $a_\v$, which is the same for all $f(X^m)$.
    Thus we have $f(X^{p_1\cdots p_k}) \in \Ann(c)$.
\qed}

\env{lemma}{
    \label{lem-polynomial-from-radical}
    Let $c$ be a finitary integral configuration and $f = \sum a_\v X^\v$ a non-trivial integer polynomial annihilator. Let $S = \makeset{\v \in \Z^d}{a_\v \ne 0}$ and define
    \env{align*}{
        g(X) = x_1\cdots x_d\prod_{\substack{\v\in S\\ \v \ne \vec{v_0}}}\left(X^{r\v}-X^{r\vec{v_0}}\right)
    }
    where $r$ is the integer from \autoref{lem-expanding-annihilator} and $\vec{v_0} \in S$ arbitrary. Then $g(Z) = 0$ for any common root $Z \in \C^d$ of $\Ann(c)$.
}
\env{proof}{
    Fix $Z$. If any of its complex coordinates is zero then clearly $g(Z) = 0$. Assume therefore that all coordinates of $Z$ are non-zero.

    Let us define for $\alpha \in \C$
    \env{align*}{
        S_\alpha &= \makesetbig{\v \in S}{Z^{r\v} = \alpha}, \\
        f_\alpha(X) &= \sum_{\v \in S_\alpha} a_\v X^\v.
    }
    Because $S$ is finite, there are only finitely many non-empty sets $S_{\alpha_1}, \dots, S_{\alpha_m}$ and they form a partitioning of $S$. In particular we have $f = f_{\alpha_1} + \dots + f_{\alpha_m}$.

    Numbers of the form $1+ir$ are relatively prime to $r$ for all non-negative integers $i$, therefore by \autoref{lem-expanding-annihilator}, $f(X^{1+ir}) \in \Ann(c)$. Plugging in $Z$ we obtain $f(Z^{1+ir}) = 0$. Now compute:
    \env{align*}{
        f_\alpha(Z^{1+ir}) &= \sum_{\v \in S_\alpha} a_\v Z^{(1+ir)\v} = \sum_{\v \in S_\alpha} a_\v Z^\v \alpha^i = f_\alpha(Z)\alpha^i}
    Summing over $\alpha=\alpha_1,\dots, \alpha_m$ gives
     \env{align*}{
        0 = f(Z^{1+ir}) &= f_{\alpha_1}(Z)\alpha_1^i + \dots + f_{\alpha_m}(Z)\alpha_m^i.
    }
    Let us rewrite the last equation as a statement about orthogonality of two vectors in $\C^m$:
    \env{align*}{
        \left( f_{\alpha_1}(Z), \dots, f_{\alpha_m}(Z) \right) \perp (\alpha_1^i, \dots, \alpha_m^i)
    }
    By Vandermode determinant, for $i \in \{0,\dots,m-1\}$ the vectors on the right side span the whole $\C^m$. Therefore the left side must be the zero vector, and especially for $\alpha$ such that $\vec{v_0} \in S_\alpha$ we have
    \env{align*}{
        0 = f_{\alpha}(Z) = \sum_{\v \in S_{\alpha}} a_\v Z^\v.
    }
    Because $Z$ does not have zero coordinates, each term on the right hand side is non-zero. But the sum is zero, therefore there are at least two vectors $\vec{v_0},\v \in S_\alpha$. From the definition of $S_\alpha$ we have $Z^{r\v} = Z^{r\vec{v_0}} = \alpha$, so $Z$ is a root of $X^{r\v}-X^{r\vec{v_0}}$.
\qed}

\env{theorem}{
    \label{thm-product-of-differences}
    Let $c$ be a finitary integral configuration with a non-trivial annihilator. Then there are non-zero $\vec{v_1}, \dots, \vec{v_m} \in \Z^d$ such that the Laurent polynomial
    \env{align*}{
        (X^{\vec{v_1}}-1)\cdots (X^{\vec{v_m}}-1)
    }
    annihilates $c$.
}
\env{proof}{
    This is an easy corollary of \autoref{lem-polynomial-from-radical}. The polynomial $g(X)$ vanishes on all common roots of $\Ann(c)$, therefore by Hilbert's Nullstellensatz there is $n$ such that $g^n(X) \in \Ann(c)$. Note that any monomial multiple of an annihilator is again an annihilator. Therefore also
    \env{align*}{
        \frac{g^n(X)}{x_1^n \cdots x_d^n X^{nr\vec{v_0}(\abs S-1)}}
    }
    is, and it is a Laurent polynomial of the desired form.
\qed}

Multiplying a configuration by $(X^\v-1)$ can be seen as a \emph{"difference operator"} on the configuration. \autoref{thm-product-of-differences} then says, that there is a sequence of difference operators which annihilates the configuration. We can reverse the process: let us start by the zero configuration and step by step \emph{"integrate"} until we obtain the original configuration. This idea gives the Decomposition theorem:

\envparam[Decomposition theorem]{theorem}{
    \label{thm-decomposition-theorem}
    Let $c$ be a finitary integral configuration with a non-trivial annihilator. Then there exist periodic integral configurations $c_1, \dots, c_m$ such that $c = c_1 + \dots + c_m$.
}

\begin{example}
\label{ex:3d_again}
Recall the 3D counter example in \autoref{ex:3d}. It is the sum
$c_1+c_2$
where $c_1(i,0,0)=1$ and $c_2(0,i,n)=1$ for all $i\in \Z$, and all other entries
are $0$. Configurations $c_1$ and $c_2$ are $(1,0,0)$- and $(0,1,0)$-periodic, respectively, so that
$(X^{(1,0,0)}-1)(X^{(0,1,0)}-1)$ annihilates $c=c_1+c_2$.\qed
\end{example}

\begin{example}
\label{ex:unbounded}
The periodic configurations $c_1,\dots ,c_m$ in \autoref{thm-decomposition-theorem} may,
for some configurations $c$,
be necessarily non-finitary. Let $\alpha\in \R$ be irrational, and define three periodic
two-dimensional configurations $c_1, c_2$ and $c_3$ by
$$
c_1(i,j)=\lfloor i\alpha\rfloor, \hspace*{1cm}
c_2(i,j)=\lfloor j\alpha\rfloor, \hspace*{1cm}
c_3(i,j)=\lfloor (i+j)\alpha\rfloor.
$$
Then $c=c_3-c_1-c_2$ is a finitary integral configuration (over alphabet $\{0,1\}$),
annihilated by the polynomial $(X^{(1,0)}-1)(X^{(0,1)}-1)(X^{(1,-1)}-1)$, but it cannot be
expressed as a sum of finitary periodic configurations.\qed
\end{example}

\section{Structure of the Annihilator Ideal}
\label{sec-ann-structure}

In the rest of the paper we focus on two-dimensional configurations.
We analyze $\Ann(c)$ using tools of algebraic geometry and provide a description of a polynomial $\phi$ which divides every annihilator. Moreover we show a theoretical result that $\Ann(c)$ is a radical ideal, which allows us to provide a stricter version of the Decomposition theorem for two-dimensional configurations.

The key ingredient needed for further analysis is the concept of a \emph{line polynomial}. Let the \emph{support} of a Laurent polynomial $f = \sum a_\v X^\v$ be defined as
\env{align*}{
    \supp(f) = \makeset{\v \in \Z^d}{a_\v \ne 0}.
}
We say that $f$ is a \emph{line Laurent polynomial} if the support contains at least two points and all the points lie on a single line. Let us call a vector $\v \in \Z^d$ \emph{primitive} if its coordinates don't have a common non-trivial integer factor. Then every line Laurent polynomial can be expressed as
\env{align*}{
    f(X) = X^{\v'}(a_n X^{n\v} + \dots + a_1 X^\v + a_0)
}
for some $a_i \in \C$, $n\geq 1$, $a_n \ne 0 \ne a_0$, $\v',\v \in \Z^d$, $\v$ primitive. Moreover,
the vector $\v$ is determined uniquely up to the sign. We define the \emph{direction} of a line Laurent polynomial to be the vector space $\gen \v \subset \Q^d$.

To simplify the notation, we prefer to write $\C[x,y]$ in the place of $\C[x_1,x_2]$.
We begin by a sequence of lemmas with a result from algebra. Recall that an ideal $A$ is \emph{prime} whenever $ab \in A$ implies $a \in A$ or $b \in A$. An ideal is \emph{radical} if $a^n \in A$ implies $a \in A$. 

\env{lemma}{\
    \label{lem-structure-of-cxy}
    \env{enumerate}{
        \item Prime ideals in $\C[x,y]$ are maximal ideals, principal ideals generated by irreducible polynomials, and the zero ideal.
        \item Every radical ideal $A \leq \C[x,y]$ can be uniquely written as a finite intersection of prime ideals $P_1, \dots, P_k$ where $P_i \not \subset P_j$ for $i \ne j$. Moreover
        \env{align*}{
            A = \bigcap_{i = 1}^k P_i = \prod_{i = 1}^k P_i.
        }
    }
}

\env{lemma}{
    \label{lem-ann-is-radical}
    Let $c$ be a two-dimensional, finitary and integral configuration with a non-trivial annihilator.
    Then $\Ann(c)$ is radical.
}

Our proof of \autoref{lem-ann-is-radical} relies on the decomposition
of two-dimensional radical ideals
into a product of primes from \autoref{lem-structure-of-cxy}, which fails in higher dimensions. However, we conjecture that \autoref{lem-ann-is-radical} is true for higher dimensions as well.

\env{lemma}{
    \label{lem-ann-structure}
    Let $c$ be as in \autoref{lem-ann-is-radical}.
    Then there exist polynomials $\phi_1, \dots, \phi_m$ and an ideal $H \leq \C[x,y]$ such that
    \env{align*}{
        \Ann(c) = \phi_1 \cdots \phi_m H
    }
    where $\phi_i$ are line polynomials in pairwise distinct directions, and $H$ is either an intersection of finitely many maximal ideals or $H = \C[x,y]$.

    Moreover $H$ is determined uniquely and $\phi_i$ are determined uniquely up to the order and multiplication by a constant.
}

Note that $H = \C[x,y]$ is not really a special case -- it covers the case when $H$ is the empty intersection.
Let us denote the number $m$ from \autoref{lem-ann-structure} by $\opc(c)$.
It is an important invariant of the configuration which provides information about its periodicity.
A two-dimensional configuration is \emph{doubly periodic} if there are two linearly independent vectors in which it is periodic. A configuration which is periodic but not doubly periodic is called \emph{one-periodic}.

\envparam[Strong decomposition theorem]{theorem}{
    \label{thm-strong-decomposition}
    Let $c$, $m=\opc(c)$, and $\Ann(c) = \phi_1 \cdots \phi_m H$ be as in \autoref{lem-ann-structure}. Let $\phi = \phi_1 \cdots \phi_m$. Then there exist configurations $c_\phi, c_H, c_1, \dots, c_m$ such that
    \env{align*}{
        c &= c_\phi + c_H \\
        c_\phi &= c_1 + \dots + c_m,
    }
    where $\Ann(c_\phi) = \gen\phi$, $\Ann(c_H) = H$ and $\Ann(c_i) = \gen{\phi_i}$. Moreover $c_\phi$ and $c_H$ are determined uniquely. Each $c_i$ is one-periodic in the direction of $\phi_i$, and $c_H$ is doubly periodic.
}

\env{corollary}{
    \label{cor-num-of-1per-directions}
    Let $c$ be as in Theorem~\ref{thm-strong-decomposition}. Then
    \env{itemize}{
        \item if $\opc(c)=0$ the configuration is doubly periodic,
        \item if $\opc(c)=1$ the configuration is one-periodic,
        \item if $\opc(c)\geq 2$ the configuration is non-periodic.
    }
}


\section{Approaching Nivat's Conjecture}
\label{sec-results}

We already know that if a finitary integral configuration $c$ satisfies the condition $P_c(m,n) \leq mn$ for some positive integers $m,n$, then it has an annihilating polynomial. The Nivat's conjecture claims that such a configuration is periodic, that is, $\opc(c) \leq 1$. Our approach is the contrapositive: assume
that $c$ is a finitary integral configuration which is non-periodic, that is, $\opc(c) \geq 2$. If $c$ does not have an annihilating polynomial, we have $P_c(m,n) > mn$ for all $m$ and $n$, and we are done.
So we assume $c$ has an annihilating polynomial so that the theory developed so far applies to $c$. We want to prove that $c$ has high local complexity.

Assuming $\opc(c) \geq 2$, let $\varphi_1$ and $\varphi_2$ be irreducible factors of $\phi_1$ and $\phi_2$.
Any annihilator of $c$ has a factor $f\in \Ann(c)$ that
can be written as $f = \varphi_1 \varphi_2 f'$ such that
$c_1=\varphi_2f'c$ and $c_2=\varphi_1f'c$ are one-periodic configurations in different directions.
Moreover, a block in $c$ determines smaller blocks in $\varphi_2f'c$ and $\varphi_1f'c$ because the multiplication by a polynomial is a local operation on the
configurations. We next estimate the number of distinct blocks in one-periodic configurations in order to lower bound the number of slightly bigger blocks in $c$.


\subsection*{Complexity of One-periodic Configurations}
\label{sec-1d-complexity}

Recall that for a finite domain $D \subset \Z^d$ we denote by $c_{\v+D}$ the
pattern extracted from the position $\v \in \Z^d$ in $c$.
Let us define a \emph{line of $D$-patterns in direction $\u \in \Z^d$} to be a set of the form
\env{align*}{
    \L = \makesetbig{c_{\v+k\u+D}}{k \in \Z}
}
for some vector $\v \in \Z^d$.

It is easy to characterize irreducible factors of line polynomials -- every line polynomial can be decomposed as
\env{align*}{
    f(X) &= X^{\v'}(a_n X^{n\v} + \dots + a_1 X^\v + a_0) \\
    &= a_n X^{\v'}(X^\v - \lambda_1)\dots(X^\v - \lambda_n)
}
where $a_0 \ne 0 \ne a_n$, $\v$ is a primitive vector and $\lambda_1, \dots, \lambda_n$ are complex roots of the polynomial $a_nt^n + \dots + a_1t + a_0$. A Laurent polynomial of the form $X^\v - \lambda_i$ is irreducible. Therefore an irreducible polynomial factor of $f$ either divides $X^{\v'}$, or has to be up to a multiplicative constant of the form $X^{\v''}(X^\v-\lambda_i)$ for some $\v'' \in \Z^d$.


The following two lemmas will be applied later on the
one-periodic configurations $c_1=\varphi_2f'c$ and $c_2=\varphi_1f'c$, respectively.
For a vector $\v = (v_1, v_2) \in \Z^2$ let us denote the size of a minimal rectangle that contains it by $Box(\v) := (|v_1|,|v_2|) \in \Z^2$.

\env{lemma}{
    \label{lem-disjoint-lines}
    Let $c$ be a two-dimensional one-periodic configuration and $\v',\v \in \Z^2$, $0 \ne \lambda \in \C$ such that $\Ann(c) = \gen{X^{\v'}(X^\v-\lambda)}$. Let $(m,n) = \bbox(\v)$. Then
    for any non-negative integers $M,N$ there are at least $Mn + mN + mn$ disjoint lines of blocks $(M+m) \times (N+n)$ in $c$ in the direction of $\v$.
}

\env{lemma}{
    \label{lem-complex-lines}
    Let $c$ be a two-dimensional one-periodic configuration and $\u',\u \in \Z^2$, $0 \ne \lambda \in \C$ such that $\Ann(c) = \gen{X^{\u'}(X^\u-\lambda)}$. Let $(m,n) = \bbox(\u)$ and $\v \in \Z^2$ be a vector in a different direction than $\u$.

    If $\L$ is any line of blocks $(M+m) \times (N+n)$ in direction $\v$ in $c$, then
    \env{align*}{
        \abs {\L} > \frac{Mn + mN}{S},
    }
    where $S$ is the positive area of the parallelogram specified by vectors $\u$ and $\v$.
}


\subsection*{Putting Things Together}

Applying Lemmas~\ref{lem-disjoint-lines} and \ref{lem-complex-lines} on the configurations
$c_1=\varphi_2f'c$ and $c_2=\varphi_1f'c$ provides the following lower bound for their common pre-image
$c'=f'c$.

\env{lemma}{
    \label{lem-two-directions-complexity}
    Let $c'$ be a two-dimensional configuration such that
    \env{align*}{
        \Ann(c') = \gen{X^{\v'}(X^{\vec{v_1}}-\lambda_1)(X^{\vec{v_2}}-\lambda_2)}
    }
    where $\lambda_1, \lambda_2 \in \C$ are non-zero, $\v', \vec{v_1}, \vec{v_2} \in \Z^2$ and $\vec{v_1}, \vec{v_2}$ are primitive vectors. Denote $(m_i,n_i) = \bbox(\vec{v_i})$. Then
    \env{align*}{
        P_{c'}(M+m_1+m_2,N+n_1+n_2) > \frac{(Mn_1+m_1N)(Mn_2+m_2N)}{m_1n_2+m_2n_1}
    }
    for all non-negative integers $M$ and $N$.
}

Let $f$ be a Laurent polynomial in two variables and $S$ its support. Let us extend the definition of the bounding box $\bbox(\cdot)$ by setting
\vspace*{-5.5mm}

\env{align*}{
    \bbox(f) = (\max_{(a,b) \in S}a - \min_{(a,b) \in S}a, \max_{(a,b) \in S}b - \min_{(a,b) \in S}b).
}

\env{corollary}{
    \label{cor-complexity}
    Let $c$ be a two-dimensional non-periodic finitary integral configuration and $f$ its annihilator. Denote $(m,n) = \bbox(f)$ and let $M \geq m, N \geq n$ be integers. Then:
    \env{enumerate}{
        \item[(a)] $P_c(M,N) > (M-m)(N-n)$.
        \item[(b)] If in the decomposition $\Ann(c) = \phi_1 \cdots \phi_{\opc(c)}H$ there are two $\phi_i, \phi_j$ such that their directions are not horizontal or vertical, then $\exists \alpha>1$:
        $$P_c(M,N) > \alpha (M-m)(N-n).$$
        \item[(c)] If $\opc(c) \geq 3$ then $$P_c(M,N) > 2(M-m)(N-n).$$
    }

}

\subsection*{The Main Result}

\env{theorem}{
    \label{thm-main-result}
    Let $c$ be a two-dimensional non-periodic configuration. Then
$P_c(M,N) > MN$
    holds for all but finitely many choices $M,N \in \N$.
}

\env{corollary}{
    If $c$ is a two-dimensional configuration such that $P_c(M,N) \leq MN$
    holds for infinitely many pairs $M,N \in \N$, then $c$ is periodic.
}

The proof (details omitted) is structured as follows. Let $c$ be non-periodic with a non-trivial annihilator,
and let $\Ann(c) = \phi H$ be the decomposition of the annihilator as in \autoref{thm-strong-decomposition}, where $\phi=\phi_1 \cdots \phi_{\opc(c)}$. We consider different ranges of $M$ and $N$.
\smallskip

\noindent
{\bf Very thin blocks.} Suppose $N$ or $M$ is so small that the support of $\phi$ does not fit
inside the $M\times N$ rectangle. Then no annihilator of $c$ fits inside the rectangle, and as in
Lemma~\ref{lem-low-cplx-has-annihilator} we see that $P_c(M,N) > MN$, no matter how large the other dimension of the rectangle is.
\smallskip

\noindent
{\bf Thin blocks.} Consider fixed $N$, large enough so that the support of $\phi$ fits inside a strip of height $N$. It can be shown 
that there exists $M_0$ such that
for all $M>M_0$ we have  $P_c(M,N) > MN$. Analogously for a fixed $M$.
\smallskip

\noindent
{\bf Fat blocks.} We prove that there are constants $M_0$ and $N_0$ such that for $M>M_0$
and $N>N_0$ we have $P_c(M,N) > MN$. This follows directly from \autoref{cor-complexity}(c) and (b), respectively, in the cases when $\opc(c)\geq 3$,  or when $\opc(c)=2$ and $\phi_1$ and $\phi_2$ are not horizontal or vertical. The cases when $\opc(c)=2$ and $\phi_1$ is vertical (or the symmetric cases) require more careful analysis. In particular, we use the observation that it is enough to consider two letter configurations:

\begin{lemma}
In any non-periodic configuration $c\in A^{\Z^2}$, letters can be merged to obtain a non-periodic
configuration $c'\in\{0,1\}^{\Z^2}$. Then $P_{c'}(D)\leq P_{c}(D)$ for all finite $D\subseteq \Z^d$.
In particular, if Nivat's conjecture holds on binary configurations it holds in general.
\end{lemma}

It is clear that the three ranges of $M$ and $N$ above cover everything so that $P_c(M,N) \leq MN$ can hold only for a finite number of $M,N\in\N$.


\printbibliography[notcategory=excluded]



\end{document}